\documentstyle[pra,aps]{revtex}

\def\ut#1{\lower1.2ex\hbox{$\mathchar"3218$}\mkern -14mu%
          \hbox to 2ex{\hss$#1$\hss}}

\begin{document}
\draft

\title{Langevin equation of collective modes of  Bose-Einstein condensates in traps}
\author{Robert Graham}
\address{Fachbereich Physik, Universit\"at-Gesamthochschule Essen\\
45117 Essen\\ Germany}
\maketitle

\begin{abstract}
A quantum Langevin equation for the amplitudes of the collective modes in
Bose-Einstein condensate is derived. The collective modes are coupled to a thermal reservoir  of quasi-particles, whose elimination
leads to the quantum Langevin equation. The dissipation rates
  are determined
via the correlation function of  the fluctuating
force and  are evaluated in the local-density approximation for the spectrum of quasi-particles  and the Thomas-Fermi approximation for the condensate.
I take great pleasure in dedicating this paper to  Gregoire Nicolis on the occasion of his sixtieth birthday.
\end{abstract}

\vskip1pc


\vskip2pc

\section{Introduction}\label{sec:1}
The realization of Bose-Einstein condensates of very rarefied evaporatively cooled gases of alkali atoms in magnetic traps \cite{1} offers the unique possibility
to test  ab initio many-body theories in the laboratory \cite{reva}. One very fertile field has been the experimental and theoretical investigation of collective modes
of the condensates, both for zero and finite temperatures. For reviews of the experimental and theoretical work
see \cite{rev1} and \cite{rev2,grif} respectively.
As opposed to conventional superfluids like He-II \cite{griffinbook,0} in the new systems the collision-less regime
is very naturally realized. In this regime the dominant damping mechanism for collective modes is Landau-damping, whose temperature dependence in spatially homogeneous condensates in the regime $k_BT$
large compared to the chemical potential $\mu$ has first been studied by Sz\'epfalusy and Kondor \cite{10}. Recent investigations  \cite{shi}, \cite{Pitaevskii} , \cite{11},\cite{Giorgini}, though more exact,
led to similar results,  differing by a prefactor  close to 1 for the damping rate.
For condensates in traps  Landau damping of low-lying modes is
more difficult to calculate, and additional approximations are needed to cope with the fact that momentum is not conserved in a trap. The damping rate of collective modes in traps has been calculated  in \cite{12} using the local density approximation and, in addition,
 a classical approximation for the correlation function whose Fourier-transform determines the cross-section of Landau-scattering. For the isotropic breathing mode in isotropic traps the  Landau-damping  has been calculated
numerically \cite{Guilleumas} by evaluating  the coupling to a great number of discrete quasi-particle modes  and subsequently
introducing some smoothing. The quasi-continuum coupled to the collective mode under study was displayed explicitly in this work. Theories using an extension of the approach of \cite{10} via the dielectric formalism \cite{reidl1,reidl2} and
 an approach via a time-dependent mean field scheme \cite{gior} have also been given.

In the present paper a very direct approach \cite{PRL} to the dissipative equilibrium and non-equilibrium dynamics of collective modes in trapped Bose-Einstein condensates via
quantum Langevin equations is put forward. Because of
the discreteness of the mode-spectrum in traps
 the problem  is formally similar to the  quantum-optical problem
of discrete modes in a laser, for which the formulation in terms of
quantum Langevin equations has been very useful \cite{9a}.

 In the next section the  microscopic description of a trapped Bose-Einstein condensed gas is briefly set up. Then we recall the basics of the
quantum Langevin equation of a boson mode. The derivation of the quantum Langevin equation of a collective mode follows.
The damping rates are then evaluated in the
local density and the Thomas-Fermi approximation. The last section contains
a discussion of our results.

\section{Microscopic equations of motion}\label{sec:2}

The weakly interacting Bose-gas in a trap
in standard notation is described by the Hamiltonian
\begin{equation}
 \hat{H}=\int\!d^3\!x
 \hat{\psi}^+\Big\{-\frac{\hbar^2}{2m}\nabla^2+
  V(\bbox{x})-\mu+\frac{U_0}{2}\hat{\psi}^+
  \hat{\psi}\Big\}\hat{\psi}\,.
 \label{eq:H}
\end{equation}
The
presence of a Bose-Einstein condensate means that many ($N_0\gg1$)
particles occupy a single normalized mode of a macroscopic classical matter wave, determined as the mode of lowest energy of the classical Hamiltonian corresponding to eq.(\ref{eq:H}). It satisfies a classical wave equation, the Gross-Pitaevskii equation \cite{5}, which we take in an extension defined by
the so-called Popov-approximation \cite{Griffin}, including the interaction of the condensate with the density $n'$ of thermal atoms, but neglecting its interaction with the pair amplitude $\langle\hat\psi\hat\psi\rangle-\psi_0^2$
of thermal particles
\begin{equation}
 -(\hbar^2/2m)\nabla^2\psi_0+
  \left(V(\bbox{x})+U_0(N_0|\psi_0(\bbox x)|^2+2n'(\bbox x))\right)\psi_0=\mu\psi_0.
\label{eq:Ha}
\end{equation}
For given $N_0$  the chemical potential $\mu$ follows by imposing the normalization condition on $\psi_0$.

The presence of the highly
occupied condensate mode makes  the decomposition of the Heisenberg
field-operator
\begin{equation}
\hat{\psi}(\bbox{x},t)=\big(\sqrt{N_0}\psi_0(\bbox{x})+
\hat{\psi}'(\bbox{x},t)\big)\exp{(-i\mu t/\hbar)}\exp{(i\phi)}
\label{2}
\end{equation}
useful.
$\sqrt{N_0}\exp{(i\phi)}$ is the  complex amplitude of  the classical condensate mode  in equilibrium.
$\hat{\psi}'(\bbox{x},t)$ is  the
field operator for the particles outside the condensate. We
shall neglect fluctuations of the number of atoms in the condensate and also
fluctuations of the phase of the condensate, which can be shown to occur on a
time-scale much longer than the relaxation-time of the collective modes \cite{PRL}.

The Hamiltonian
splits up according to $\hat{H}=H_0+\hat{H_1}+\hat{H_2}+\hat{H_3}+\hat{H_4}$, with a c-number term $H_0$ which need not concern us here, and
\[\hat H_1=\sqrt{N_0}\int d^3x \left(\big(V(\bbox x)-\mu+U_0(N_0|\psi_0|^2+2n')\big)\psi^*_0\hat\psi'+\big(hermitian \quad conjugate\big)\right)
\]
\[\hat H_2=\int d^3x\left(\hat\psi'^+(-\frac{\hbar^2\nabla^2}{2m}+V(\bbox x)+2U_0n'-\mu)\hat\psi'+\frac{U_0N_0}{2}(\psi^{*2}_0\hat\psi'^2+\psi_0^2\hat\psi'^{+2}+4|\psi_0|^2\hat\psi'^+\hat\psi')\right)
\]
\[\hat H_3=U_0\sqrt{N_0}\int d^3x\left(\psi_0^*(\hat\psi'^+\hat\psi'-2n')\hat\psi'+(hermitian\quad conjugate)\right)
\]
\[\hat H_4=\frac{U_0}{2}\int d^3x\hat(\psi'^+\hat\psi'^+\hat\psi'\hat\psi'-4n'
\hat \psi'^+\hat \psi').
\]
The splitting is here done in such a way that the term $\hat{H_1}$ vanishes due to eq.(\ref{eq:Ha}) and  the part $\hat{H_2}$ describes the linearized quantum excitations around the solution of (\ref{eq:Ha}). $\hat{H_2}$ is diagonalized by introducing
quasi-particle operators $\hat{\alpha}_{\nu},\hat{\alpha}_{\nu}^{+}$ by
the standard Bogoliubov transformation
\[
\hat{\psi}'(\bbox{x},t)=\sum_\nu
\big(u_\nu(\bbox{x})\hat{\alpha}_\nu(t)+
v^*_\nu(\bbox{x})\hat{\alpha}_\nu^+(t)\big),
\]
where $u_\nu$, $v_\nu$ satisfy the usual Bogoliubov-De Gennes equations
\begin{equation}
 \left(\begin{array}{cc}
         -\frac{\hbar^2}{2m}\nabla^2+U_{\rm eff}(\bbox{x})-\hbar\omega_\nu &
         K(\bbox{x})\\
          K^*(\bbox{x}) &
            -\frac{\hbar^2}{2m}\nabla^2+U_{\rm eff}(\bbox{x})+\hbar\omega_\nu
             \end{array}\right)
             {u_\nu(\bbox{x}) \choose v_\nu(\bbox{x})}=0,
\label{eq:2.18}
\end{equation}
with the abbreviations
\begin{eqnarray}
 &&  U_{\rm eff}(\bbox{x})=V(\bbox{x})
  +2U_0(N_0|\psi_0(\bbox{x})|^2+n'(\bbox x))-\mu\nonumber\\
 && K(\bbox{x})=N_0U_0\psi_0(\bbox{x})^2\,.
 \label{eq:2.19}
\end{eqnarray}
Eq.~(\ref{eq:2.18}) is consistent with  the ortho-normality conditions
 $ \int d^3x(u_\nu u_\mu^*-v_\nu v_\mu^*)=\delta_{\nu \mu}$ and
$ \int d^3r(u_\nu^*v_\mu-u_\mu^*v_\nu)=0$,
which guarantee the Bose commutation relations of the $\alpha_\nu$,
$\alpha_\mu^+$.
The decomposition of $\hat{\psi}$ and $\hat{\psi}'$ together imply that
$N=N_0+\int n'(\bbox x)d^3x $
with
$n'=\sum_\mu\big(\bar n_\mu(|u_\mu|^2+|v_\mu|^2)+|v_\mu|^2\big)$.

Within Bogoliubov(-Popov) theory the terms $\hat H_3, \hat H_4$ of the total Hamiltonian are neglected and the quasi-particle operators $\hat{\alpha}_{\nu}$  in the Heisenberg-picture obey the
Heisenberg equations of motion
$\dot{\hat{\alpha}}_\nu=-i\omega_\nu\hat{\alpha}_\nu$.
In this approximation the collective modes and the quasi-particles have infinite lifetime. In reality, however, the lifetime will be limited
by the scattering of quasi-particles in any given mode $\nu$ with other quasi-particles from the thermal reservoir, which is described by  $\hat H_3$ and $\hat H_4$. One way to describe this is the quantum Langevin equation.

\section{quantum Langevin-equation of a harmonic oscillator}\label{sec:3a}

Let us recall here briefly the quantum Langevin equation, in Markoff approximation, of a harmonic oscillator as it is commonly used in quantum optics \cite{9a,Gardiner}.
For a detailed discussion of its derivation I refer to \cite {Gardiner}.
Eq.(3.4.63) of that reference states the quantum Langevin equation
in resonance or 'rotating wave' approximation for a harmonic oscillator,  described by the Bose operators $\hat a, \hat a^+$, in interaction with a thermal reservoir at temperature $T$. It takes the form
\begin{equation}
\dot{\hat a}(t)=-i\Omega \hat{a}(t)-\gamma \hat{a}(t) + \hat{\xi}(t)
\label{Langevin}\end{equation}
where $\Omega$ is the frequency of the oscillator  including a frequency shift due to the oscillator's coupling to a heat reservoir, $\gamma$ is the damping rate, and $\hat{\xi}(t)$ is a Gaussian noise-operator.
In Markoff approximation it has the correlation functions
\begin{equation}
\langle\hat\xi^{+}(t)\hat\xi(t')\rangle=\frac{2\gamma}{\exp(\hbar\Omega/k_BT)-1}\delta(t-t')
\label{FD1}\end{equation}
ensuring the correct normally ordered expectation values in
equilibrium, and
\begin{equation}
\langle[\hat\xi(t),\hat\xi^+(t')]\rangle=2\gamma\delta(t-t')
\label{FD2}\end{equation}
ensuring the correct Bose commutation relations of $\hat a(t), \hat a^+(t)$ for all times. The fluctuation-dissipation relation therefore permits us to infer
the properties of $\hat\xi(t)$ if the coefficient of the dissipative term is known. Alternatively we can infer the dissipation rate $\gamma$ from a microscopic expression for $\hat\xi(t)$ either by using (\ref{FD1}) or, alternatively, (\ref{FD2}).

\section{quantum Langevin-equation of collective modes}\label{sec:4}

We shall here
confine our attention to the dynamics of the low-lying collective modes in the
collision-less regime.
The interaction of the collective modes with the thermal quasi-particles
is described by the Hamiltonian $\hat H_3+\hat{H}_4$
 not yet contained in  the Bogoliubov(-Popov) approximation.
Because it contains the large factor $\sqrt{N_0}$ the
contribution $\hat H_3$ dominates over $\hat H_4$ and the latter can be neglected in the following.
Inserting the Bogoliubov transformation in $\hat{H}_3$  and going to the interaction picture  with respect to the unperturbed Bogoliubov-Popov Hamiltonian,
  $\hat{\tilde H}_3$  in interaction representation  takes the form
\begin{eqnarray}
\hat{\tilde H}_3=
 \frac{\sqrt{N_0}}{2}\sum_{\kappa\nu\mu}\big\{
&&(M_{\kappa,\nu\mu}^{(1)}+(M_{\nu\mu,\kappa}^{(2)})^*)\hat{\alpha}_\kappa^+\hat{\alpha}_\nu\hat{\alpha}_\mu\exp[i(\omega_\kappa-\omega_\nu-\omega_\mu)t]\nonumber\\
&&+
(hermitian \quad conjugate)\big\}
+(nonresonant \quad terms).
 \end{eqnarray}
where $\bar n_\mu=1/(\exp(\hbar\omega_\mu/k_BT)-1)$ is the thermal number
of quasi-particles at frequency $\omega_\mu$.

Nonresonant terms,  in which the frequencies of the quasi-particles cannot add up to zero, have not been written out explicitly, because
later-on we shall restrict ourselves to the resonance or rotating wave approximation in
which they don't contribute. The relevant matrix elements $M^{(1)},M^{(2)}$ are
\begin{eqnarray}\label{M}
M^{(1)}_{\kappa,\nu\mu}=&&2U_0\int d^3x\psi_0v_\nu(u^*_\kappa
u_\mu+\frac{1}{2}v^*_\kappa v_\mu)+(\nu \leftrightarrow\mu)\nonumber\\
 M^{(2)}_{\nu\mu,\kappa}=&&2U_0\int d^3x\psi_0u^*_\nu (v^*_\mu
v_\kappa +\frac{1}{2}u^*_\mu u_\kappa)+(\nu \leftrightarrow\mu).
 \end{eqnarray}
$M^{(1)}_{\kappa,\nu\mu}$ describes a scattering process
in which one atom is scattered out of the condensate by the absorption
of the two quasi-particles $\nu, \mu$ out of  and the emission
of the new quasi-particle $\kappa$ into  the thermal bath. Likewise
$M^{(2)}_{\nu\mu,\kappa}$ describes a scattering process where an incoming thermalquasiparticle
$\kappa$ is absorbed, again an atom is kicked out of the condensate, and two
quasi-particles $\nu, \mu$ are emitted into the thermal bath. The scattering amplitudes for both processes are linearly superposed due to the phase-coherence of the condensate on the time-scale of the relaxation process induced by the scattering process.

Taking $\hat{\tilde H}_3$ into account the equations of motion of $\hat{\tilde\alpha}_\nu(t)$ in the interaction picture $\hat{\alpha}_\nu(t)=\hat{\tilde\alpha}_\nu(t)\exp(-i\omega_\nu t)$   become
\begin{eqnarray}
\dot{\hat{\tilde\alpha}}_\nu=-\frac{i}{\hbar}\frac{\sqrt{N_0}}{2}\sum_{\kappa\mu}\big\{
&&\big[M^{(1)}_{\nu,\kappa\mu}+(M^{(2)}_{\kappa\mu,\nu})^*\big]
\hat{\tilde\alpha}_\kappa\hat{\tilde\alpha}_\mu\exp[i(\omega_\nu-\omega_\mu-\omega_\kappa)t]\nonumber\\
+2&&\big[(M^{(1)}_{\kappa,\nu\mu})^*+M^{(2)}_{\nu\mu,\kappa}\big]
\hat{\tilde\alpha}_\mu^+\hat{\tilde\alpha}_\kappa \exp[i(\omega_\nu+\omega_\mu-\omega_\kappa)t]\big\}\label{Heis}.
\end{eqnarray}
If the back-action of the collective mode on the quasi-particle operators $\hat \alpha_\mu, \hat \alpha_\kappa$ in (\ref{Heis}) can be ignored,
the new  term in this equation of motion acts like an effective random force operator. In the resonance approximation the average of this force operator vanishes. In addition it is white noise, in good approximation, if the frequencies $\omega_\kappa-\omega_\mu-\omega_\nu$ and $\omega_\kappa+\omega_\mu-\omega_\nu$ it contains form
a  closely spaced quasi-continuum  near  $0$ in a  neighborhood which is broad compared to the resulting damping rate $\gamma_\nu$.
For an explicit display of this quasi-continuum in a concrete
example see \cite{Guilleumas}. In as much as this condition
is satisfied for large condensates the Markoff-assumption made earlier is justified. All terms in the fluctuating force term not containing frequencies near frequency $0$ are non-resonant and can be omitted in comparison with resonant terms.

As we recalled in the previous section,  the noise term is
always accompanied by a dissipative term, and, due to the Kramers-Kronig relation, also by a frequency shift. Thus the complete quantum Langevin equation in resonance approximation and in Markoff approximation must take the form of (\ref{Langevin})
\begin{equation}
\dot{\hat\alpha}_\nu=-i(\omega_\nu+\delta_\nu)\hat\alpha_\nu-\gamma_\nu\hat\alpha_\nu+\hat\xi_\nu(t)
\end{equation}
where $\hat\xi_\nu(t)$ is given by the second term in (\ref{Heis}).

The damping rates $\gamma_\nu$
will be derived below, but we can also simply use (\ref{FD2}) and represent them in the form
\begin{eqnarray}
\gamma_\nu=\frac{1}{2}\int_{-\infty}^{+\infty}dt\langle[\hat\xi_\nu(t),\hat\xi_\nu^+(0)]\rangle .\label{previous}
\end{eqnarray}
Evaluating the commutator, taking the thermal expectation value, and performing the time integral in (\ref{previous}) we obtain
\begin{eqnarray}
\gamma_\nu=\frac{\pi N_0}{\hbar^2}\sum_{\kappa,\mu}\big \{&&|(M_{\kappa,\nu\mu}^{(1)})^*+M_{\nu\mu,\kappa}^{(2)}|^2(\bar{n}_\mu-\bar{n}_\kappa)\delta(\omega_\kappa-\omega_\mu-\omega_\nu)\nonumber\\
&&+|M_{\nu,\kappa\mu}^{(1)}+(M_{\kappa\mu,\nu}^{(2)})^*|^2(\bar{n}_\kappa+\frac{1}{2})\delta(\omega_\kappa+\omega_\mu-\omega_\nu)\big\}
\label{gammanu}
\end{eqnarray}
The first term describes Landau-damping of the mode $\nu$ by scattering a quasi-particle from mode $\mu$ to mode $\kappa$ and is equivalent to a result derived in \cite{Pitaevskii} by the golden rule.
The second term in eq.(\ref{gammanu}) describes Beliaev damping, where the mode $\nu$ decays into two modes $\kappa, \mu$. It survives even for $T\rightarrow 0$ where $\bar{n}_\kappa \rightarrow 0$ for all modes. However, for low-lying modes in traps there are only very few modes, or no modes at all, into which decay under energy conservation can occur, and  this contribution to the damping is then negligible.

Let us now {\it derive} the dissipative term of the quantum Langevin equation. To this end we consider the equations of motion
for $
\frac{d}{dt}(\hat{\tilde\alpha}_\mu^+\hat{\tilde\alpha}_\kappa)$ and $\frac{d}{dt}(\hat{\tilde\alpha}_\mu\hat{\tilde\alpha}_\kappa)$, keeping
again only the resonant terms. Integrating these equations over time
from $-\infty$ to $t$ and inserting the result back into the equation
of motion for $\hat{\tilde\alpha}_\nu$ we obtain
\begin{eqnarray}
\frac{d}{dt}\hat{\tilde\alpha}_\nu=
-\hat{\tilde\alpha}_\nu\frac{\sqrt{N_0}}{\hbar^2}\sum_{\mu\kappa}\left(\frac{(\bar n_\mu+1/2)|M_{\nu,\kappa\mu}^{(1)}+(M_{\kappa\mu,\nu}^{(2)})^*|^2}{\epsilon+i(\omega_\mu+\omega_\kappa-\omega_\nu)}+\frac{(\bar n_\mu-\bar n_\kappa)|(M_{\kappa,\nu\mu}^{(1)})^*+M_{\nu\mu,\kappa}^{(2)}|^2}{ \epsilon+i(\omega_\kappa-\omega_\mu-\omega_\nu)}\right)\nonumber\\
-\frac{i}{\hbar}\frac{\sqrt{N_0}}{2}\sum_{\kappa\mu}\big\{
\big[M^{(1)}_{\nu,\kappa\mu}+(M^{(2)}_{\kappa\mu,\nu})^*\big]
\hat{\tilde\alpha}_\kappa(-\infty)\hat{\tilde\alpha}_\mu(-\infty)\exp[i(\omega_\nu-\omega_\mu-\omega_\kappa)t]\\
+2\big[(M^{(1)}_{\kappa,\nu\mu})^*+M^{(2)}_{\nu\mu,\kappa}\big]
\hat{\tilde\alpha}_\mu^+(-\infty)\hat{\tilde\alpha}_\kappa(-\infty) \exp[i(\omega_\nu+\omega_\mu-\omega_\kappa)t]\big\}\nonumber
\end{eqnarray}
where the limit $\epsilon\rightarrow +0$ is implied.
The second term on the right hand side is the fluctuating force term again, now more rigorously expressed in terms of the reservoir operators at the initial time at $-\infty$.
Taking the limit with $(\epsilon-i\omega)^{-1}\rightarrow \pi\delta(\omega)+iP/\omega$, where $P/\omega$ denotes the principal part
under a frequency integral, we obtain the result (\ref{gammanu}) for the damping rate and also the frequency shifts $\delta_\nu$ in the quantum Langevin equation. They are given by the Kramers-Kronig relation
\begin{equation}
\delta_\nu=-\frac{1}{\pi}P\int d\omega\frac{\gamma(\omega)}{\omega-\omega_\nu}\label{deltanu}
\end{equation}
where we defined   $\gamma(\omega_\nu)= \gamma_\nu$.

\section{Damping rates}\label{sec:4a}
 In the following we shall neglect the second term in (\ref{gammanu}), because as discussed it cannot contribute for low lying modes. Our goal in this section is the evaluation of the first term in (\ref{gammanu}) in a well defined approximation, the local density  and the Thomas-Fermi approximation.
The local density approximation amounts to the treatment of the quasi-continuum of the spectrum of frequencies $\omega_\kappa-\omega_\mu-\omega_\nu$ as a continuum whose density is given by the semiclassical mode-densities of the
frequencies $\omega_\mu, \omega_\kappa$. Why these frequencies lie much higher than the collective mode frequency $\omega_\nu$ will become clear below.
The Thomas-Fermi approximation applies to large condensates \cite{Peth} and amounts to neglecting the kinetic energy term in the Gross-Pitaevskii equation.
The collective modes satisfy $E_\nu\ll\mu=U_0|\psi_0(0)|^2$ and
can be represented as \cite{rev2}
\begin{eqnarray}
u_\nu (\bbox x)= && \left( \sqrt\frac{U_0 n_0 (\bbox x)}{2\hbar \omega_\nu}+
 \frac{1}{2}\sqrt\frac{\hbar\omega_\nu}{2U_0n_0(\bbox x)}\right)
\chi_\nu(\bbox x)\nonumber\\
 \label{eq:L}\\
v_\nu(\bbox x)=
 && \left(-\sqrt\frac{U_0n_0(\bbox x)} {2\hbar\omega_\nu}+
 \frac{1}{2} \sqrt\frac{\hbar\omega_\nu}{2U_0n_0(\bbox x)}\right)\chi_\nu (\bbox x)\nonumber
\end{eqnarray}
with
       $\int\!\!d^3 x | \chi_\nu(\bbox x) |^2 = 1$.
The functions of the low-lying states $\chi_\nu (\bbox x)$ are known
\cite{stringari,fetter,ob,fliesser,csordas}
in the hydrodynamic (long-wavelength) and Thomas-Fermi approximation (and neglecting the influence of the thermal cloud which sits mainly outside the condensate and therefore has little influence on its collective excitations).   In spatially isotropic parabolic traps they have the form
\cite{stringari}
\begin{equation}
\chi_\nu(\bbox x)=\frac{1}{r_{TF}^{3/2}}P_{\ell_\nu}^{(2n_\nu)}(x/r_{TF})(x/r_{TF})^\ell_\nu Y_{\ell_\nu m_\nu}(\theta,\varphi)\Theta(1-x/r_{TF})
\end{equation}
The normalized polynomials  $P_\ell^{(2n)}(x)$ are known explicitly \cite{fetter}.

The high-lying quasiparticle modes can be represented as \cite{rev2}
\begin{equation}
u_\kappa(\bbox x)= \frac{E_\kappa+p^2_\kappa/2m}{\sqrt{2E_\kappa p^2_\kappa/m}}e^{i\bbox{p}_\kappa\cdot\bbox x/\hbar}, \qquad
v_\kappa (\bbox x)= - \frac{E_\kappa-p^2_\kappa/2m}{\sqrt{2E_\kappa p^2_\kappa/m}}e^{i\bbox{p}_\kappa\cdot\bbox x/\hbar}\label{N}
\end{equation}
with the local energies in Thomas-Fermi approximation
\begin{equation}
    E_\kappa=   E(p_\kappa,\bbox x) = \sqrt{(\frac{p_\kappa^2}{2m}
                  + |U_0n_0(\bbox x)|)^2-U_0^2n_0^2(\bbox x)\Theta(\mu-V(\bbox x))}
\end{equation}
and $n_0(\bbox x)=N_0|\psi_0(\bbox x)|^2=(\mu/U_0)(1-\sum_i (x_i/r_{TF}^{(i)})^2)$, and $r_{TF}^{(i)}=\sqrt{2\mu/\omega_i^2}$
are the three main Thomas-Fermi radii.

Let us consider now the Landau-damping of a low-lying phonon mode $\omega_\nu$.
If the modes $\mu,\kappa$ involved in the scattering process were also low-lying we could use eq.(\ref{eq:L})  and would obtain,
with $E_\kappa=E_\nu+E_\mu$
\begin{equation}
       (M^{(1)}_{\kappa,\nu\mu})^* + M^{(2)}_{\nu \mu,\kappa}
       =
       \frac{3U_0}{4\sqrt2} \int\!\!d^3 x \psi_0 \chi_\kappa \chi^*_\mu \chi^*_\nu             \sqrt{\frac{E_\mu E_\nu E_\kappa}{U^3_0n^3_0(x)}}
\end{equation}
However, in the limit of low-lying modes where $E_\nu, E_\mu, E_\kappa \ll U_0 n_0$ this matrix element becomes very small, i.\,e. low-lying modes cannot significantly contribute to Landau damping of other low-lying modes. Therefore the relevant modes $\mu, \kappa$ are in fact not low lying, local density approximation is applicable, and we can use   eq.(\ref{N}) for their representation.
The matrix-element for $E_\nu<<E_\mu,E_\kappa$ can be expanded in $E_\nu/\mu$ to lowest order around $E_\kappa=E_\mu$ and becomes then
\begin{equation}
(M^{(1)}_{\kappa,\nu\mu})^* + M^{(2)}_{\nu \mu,\kappa}=\sqrt{\frac{E_\nu U_0}{2N_0}}\int d^3x\chi^*_\nu(\bbox x)\exp(i(\bbox p_\kappa-\bbox p_\mu)\bbox\cdot\bbox x)) F(E_\mu, p_\mu)
\end{equation}
with
\begin{equation}
       F(E_\mu, p_\mu)
     =
       \frac{p^2_\mu}{2m}\frac{3E_\mu^2+(p^2_\mu/2m)^2}{E_\mu(E^2_\mu+(p^2_\mu/2m)^2}
\end{equation}
It will be very convenient later to express the product
$\chi_\nu^*(\bbox x)\chi_\nu(\bbox x')$ by the associated Wigner-function $W_\nu$ via
\begin{equation}
\chi_\nu^*(\bbox x)\chi_\nu(\bbox x')
=\int \frac{d^3k}{(2\pi)^3}e^{-i\bbox k\bbox\cdot(\bbox x-\bbox x')}W_\nu(\frac{1}{2}(\bbox x+\bbox x'),\bbox k)\label{Wigner}
\end{equation}
In the following we denote
\begin{equation}
(\bbox x+\bbox x')/2\rightarrow\bbox x \qquad \bbox x-\bbox x'\rightarrow\bbox r\nonumber
\end{equation}
The rate for Landau-damping can then be written as
\begin{eqnarray}
\gamma_\nu
=
 CE_\nu^2&&\int d^3x\int \frac{d^3k}{(2\pi)^3}W_\nu(\bbox x,\bbox k)
\sum_\mu\nonumber\\
&&\sum_\kappa F^2(E_\mu,p_\mu)\int d^3r  e^{\frac{i}{\hbar}(\bbox{p}_\kappa-\bbox{p}_\mu-\hbar \bbox k)\bbox\cdot
     \bbox{r}}
       \frac{\delta\,(E_\kappa-E_\mu-E_\nu)}
         {\sinh^2\left(\frac{E_\mu}{2k_BT}\right)}
\end{eqnarray}
with
\begin{equation}
C = \frac{\pi}{8\hbar}\frac{ U_0}{k_B T}
\end{equation}

The sums over the states $\mu$  and $\kappa $ are only symbolic, because in the local-density approximation the discrete states have been replaced by a continuum which is normalized on the $\delta$-function. Concretely, under the integral over $\bbox x$ the sums over the energy levels of the scattering quasi-particles $\mu, \kappa$ are in the local density approximation replaced locally by classical phase-space averages for fixed $E_\mu, E_\kappa$ and final integration over $E_\mu$ and $E_\kappa$ which takes automatically care of the normalization on the $\delta$-function. Thus
\begin{equation}
       \sum\nolimits_\mu ...\rightarrow \sum_\mu\nolimits^{(\bbox x)}... = \int dE_\mu \int\frac{d^3p_\mu}{(2\pi\hbar)^3}\delta \big( E_\mu - E\left( p_\mu, \bbox x\right )\big)...
\label{eq:semi}
\end{equation}

In the following $\sum_\mu$ and $\sum_\kappa$ will be interpreted according to eq.(\ref{eq:semi}).
The spatial integration over $\bbox{r}$ can be done and produces a momentum-conservation factor $\left(2\pi\hbar^3\right) \delta^{(3)} \left( \bbox{p_\kappa}-\bbox{p_\mu} -\hbar\bbox k\right)$. Next the integrations over $\bbox{p_\kappa}$ and $E_\kappa$ contained in $\sum_\kappa$ can be performed, which just cancel the $\delta$-functions of overall momentum and energy conservation and replace everywhere else $E_\kappa\rightarrow E_\mu+E_\nu$ and $\bbox{p_\kappa}\rightarrow \bbox{p_\mu}+\hbar\bbox{k}$. Then the expression for $\gamma_\nu$ is reduced to
\begin{eqnarray}
       \gamma_\nu = CE_\nu^2 &&\int d^3x\int \frac{d^3k}{(2\pi)^3} W_\nu(\bbox x,\bbox k)\int dE_\mu \nonumber\\
&&\int\frac{d^3p_\mu}{(2\pi\hbar)^3}\delta \big( E_\mu - E\left( p_\mu, \bbox x \right )\big)
\frac{F^2(E_\mu,p_\mu)}
            {\sinh^2 \left(\frac{E_\mu}{2k_BT}\right)}
       \delta\big(
       E_\mu + E_\nu - E\left(|\bbox p_\mu+\hbar\bbox k|, \bbox x\right)  \big)      \label{eq:oben}
\end{eqnarray}
  Next the integration over the {\it directions} of $\bbox{p}_\mu$ relative to $\bbox k$ is carried out by using up the second of the two $\delta$-functions  explicitly displayed in eq. (\ref{eq:oben}). This produces a factor $2\pi$ for the azimuthal angle, and a factor
$|\partial E(p_\mu, \bbox x)/\partial p^2_\mu|^{-1} (2p_\mu\hbar k)^{-1}$
 from the integration over $\cos \theta$ between -1 and 1, where $\theta$ is the angle between $\bbox k$ and $\bbox p_\mu$.
Finally the integration over
the absolute value $p_\mu$ is done using up the last $\delta$-function, which picks out the $\bbox x$-dependent momentum-value $
p^{(0)}_\mu=\sqrt{2m}\sqrt{\sqrt{E_\mu^2+U_0^2n_0^2(\bbox x)}-U_0n_0(\bbox x)}.
$ leaving us with the expression
\begin{equation}
      \gamma_\nu
    =
      \frac{CE_\nu^2}{4\pi^2\hbar^3}
      \int d^3x\int\frac{d^3k}{(2\pi)^3}\frac{W_\nu(\bbox x,\bbox k)}{\hbar k}
      \int dE_\mu
      \frac{F^2(E_\mu,p^{(0)}_\mu)}
           {4\big(\frac{\partial E(p^{(0)}_\mu,\bbox x)}{\partial (p_\mu^{(0)})^2}\big )^2\sinh^2\left(\frac{E_\mu}{2k_BT}\right)}
      \label{Z}
\end{equation}

We now have to face the difficulty to evaluate the conditional average of $(\hbar k)^{-1}$. The rigorous way to do this, which unfortunately leads to multiple integrals which are tedious to evaluate, is to invert (\ref{Wigner})
which yields
\begin{equation}
\int \frac{d^3k}{(2\pi)^3}\frac{W_\nu(\bbox x,\bbox k)}{\hbar k}=\int d^3r\frac{\chi_\nu^*(\bbox{x+r}/2)\chi_\nu(\bbox{x-r}/2)}{2\pi^2\hbar r^2}\label{1}
\end{equation}
A much simpler  way consists in expressing the desired average by
the local sound-velocity $\sqrt{\mu/m}\bar c_\nu(\bbox x)$ defined by
\begin{equation}
\int \frac{d^3k}{(2\pi)^3}\frac{W_\nu(\bbox x,\bbox k)}{\hbar k}=\sqrt{\frac{\mu}{m}}\frac{\bar c_\nu(\bbox x)}{E_\nu}|\chi_\nu(\bbox x)|^2.
\label{2a}\end{equation}
and estimating the dimensionless sound-velocity $\bar c_\nu(\bbox x)$ semi-classically as $\bar c_\nu(\bbox x)\approx\sqrt{1-(\bbox x/r_{TF})^2}$ with the geometrical
mean Thomas-Fermi radius $r_{TF} = \left(2\mu/m\bar\omega^2\right)^{\frac{1}{2}}$. Of course the use of the semi-classical approximation for the low lying collective mode is highly questionable and cannot be quantitatively accurate. Still we may like to use it as a rough estimate in a case where an accurate evaluation is not required or too time consuming. Below we shall  check  this approximation in two cases,
where it cannot be expected to be particularly good.

We now  introduce scaled  variables   $\tilde E_\mu = E_\mu/(N_0 U_0\psi_0^2(\bbox x))$ and $\tilde{\bbox x} = \bbox x/r_{TF}$, $\tilde{\bbox x}' = \bbox x'/r_{TF}$ with  dimensionless mode-functions $\tilde{\chi}_\nu(\tilde{\bbox x}) = r_{TF}^{\frac{3}{2}} \chi_\nu(\bbox x)$.
Altogether, using (\ref{1}),
we are left with the result
\begin{equation}
      \gamma_\nu
    =
      \frac{(a^3n_0(0))^{1/2}E_\nu^2}{2(2\pi)^{3/2}\hbar^2\bar\omega}H_\nu(\frac{k_BT}{\hbar})
\label{Z1}
\end{equation}
with the dimensionless function
\begin{eqnarray}
H_\nu(z)=
      \int d^3\tilde x \int d^3\tilde x'&&\frac{\tilde{\chi}_\nu^*(\tilde{\bbox x})\tilde{\chi}_\nu(\tilde{\bbox x}')(1-(\tilde{\bbox x}+\tilde{\bbox x}')^2/4)}{(\tilde{\bbox{x}}-\tilde{\bbox x}')^2}\nonumber\\
&&\cdot\frac{1}{z}\int d\tilde{E}_\mu
      \left(\frac{2\tilde{E}_\mu+1-\sqrt{\tilde E_\mu^2+1}}
    {(\tilde{E}_\mu^2+1)\sinh\big(\frac{1}{2z}\tilde{E}_\mu(1-(\tilde{\bbox x}+\tilde{\bbox x}')^2/4)\big)}\right)^2
      \label{Z2}
\end{eqnarray}
For $z>>1$ the functions $H_\nu$ become linear in $z=k_BT/\mu$ and reduce to
\begin{equation}
H_\nu(z)\asymp
    3\pi z \int d^3\tilde x \int d^3\tilde x'\frac{\tilde{\chi}_\nu^*(\tilde{\bbox x})\tilde{\chi}_\nu(\tilde{\bbox x}')}{(\tilde{\bbox{x}}-\tilde{\bbox x}')^2(1-(\tilde{\bbox x}+\tilde{\bbox x}')^2/4)}      \label{Z3}
\end{equation}

The result  for the spatially homogeneous case \cite{shi}
can be recovered from eq.(\ref{Z3}) for $k_BT>>\mu$ by using the scaled homogeneous condensate density $1-\tilde x^2\rightarrow 1$, the phonon energy $E_\nu = \sqrt{\mu/m}
\hbar k_\nu$, and  normalized plane waves
to evaluate $\int d^3\tilde x \int d^3\tilde x'\tilde{\chi}_\nu^*(\tilde{\bbox x})\tilde{\chi}_\nu(\tilde{\bbox x}')/(\tilde{\bbox{x}}-\tilde{\bbox x}')^2=(2\pi^2/k_\nu)\sqrt{m\bar\omega^2/2\mu}$  which, together with (\ref{Z1},\ref{Z3}) yields
$\gamma_\nu=\frac{3\pi}{8}ak_\nu \frac{k_BT}{\hbar}$.

For isotropic traps the asymptotic result (\ref{Z3})  becomes
\begin{eqnarray}
H_{n_\nu \ell_\nu m_\nu}(z)\asymp&&
6\pi z\frac{(2\ell_\nu+1)(\ell_\nu-m_\nu)!}{(\ell_\nu+m_\nu)!}\int\limits_{0}^{1} d \tilde x
                   \tilde x^2 P_{\ell_\nu}^{(2n_\nu)}(\tilde{x})\tilde{x}^{\ell_\nu}\int\limits_{0}^{1}
d\tilde x'\tilde x'^2
P_{\ell_\nu}^{(2n_\nu)}(\tilde{x'})\tilde{x'}^{\ell_\nu}\nonumber\\
&&\int\limits_{-1}^{1}\int\limits_{-1}^{1}
\int\limits_{0}^{2\pi}\frac{d(\cos\theta) d(\cos\theta') d\phi P_{\ell_\nu}^{m_\nu}(\cos\theta)P_{\ell_\nu}^{m_\nu}(\cos\theta')\exp(-im_\nu\phi)}{(2\tilde x\tilde x'(\cos\theta\cos\theta'
+\sin\theta\sin\theta'\cos\phi)-2)^2-(\tilde x^2+\tilde x'^2-2)^2}
\label{3}\end{eqnarray}
where the functions $P_\ell^m(cos\theta)$ are the associated Legendre functions appearing in the spherical harmonics.

If instead of (\ref{1}) we use (\ref{2a})  to evaluate the conditional average of $(\hbar k)^{-1}$ we obtain in place of (\ref{3})
\begin{eqnarray}
H_{n_\nu \ell_\nu m_\nu}(z)\asymp&&
z\frac{3\sqrt{2}\pi^3\hbar\bar\omega}{E_\nu}\int\limits_{0}^{1} d \tilde x
                   \frac{\tilde x^2}{ \sqrt{(1-\tilde x^2)}}(P_{\ell_\nu}^{(2n_\nu)}(\tilde{x})\tilde{x}^{\ell_\nu})^2\label{4a}\end{eqnarray}
As was already emphasized, this result can only serve as a rough estimate for  (\ref{3}).

In the simplest case  $n_\nu=1, \ell_\nu=0$, which is the isotropic
fundamental breathing mode, we have $P_0^0(\cos(\theta))=1, P_1^{(0)}(x)=\frac{3}{2}\sqrt{7}(1-\frac{5}{3}x^2)$. In this case
(\ref{3}) can be reduced to the numerical evaluation of a two-dimensional integral and we obtain
\begin{equation}
 \gamma_{0,0}\asymp 26.42..\omega_0(a^3n_0(0))^{1/2}\frac{k_BT}{\mu}.
\label{p}\end{equation}
Eq.(\ref{4a}) yields via elementary integration
$\gamma_{0,0}\asymp 27.27..\omega_0(a^3n_0(0))^{1/2}k_BT/\mu$
which agrees surprisingly well with the more rigorous result (\ref{p}). Can this be considered typical? The answer is negative:

The simple result (\ref{4a})  lends itself to further evaluation
for modes with $\ell_\nu\neq 0$. For the surface modes with $n_\nu=0, \ell_\nu\neq 0$ we obtain the estimate
\begin{equation}
 \gamma_{0,\ell_\nu}\asymp \omega_0(a^3n_0(0))^{1/2}\frac{k_BT}{\mu}\frac{3\pi^2}{4}\sqrt{\ell_\nu}\frac{\Gamma(\ell_\nu+5/2)}{\Gamma(\ell_\nu+2)}.
\end{equation}
In this case a numerical comparison with the more accurate
expression  (\ref{3}) for the case $\ell_\nu=2$
shows that the latter is about 30 percent smaller, probably giving us a realistic impression of the accuracy of the approximation for $\bar c_\nu(\bbox x)$. For larger values of $\ell_\nu$ and $n_\nu$ the accuracy of this estimate can be expected to improve.

\section{Discussion and conclusion}\label{sec:5}
In the present paper the many-body problem of collective modes in Bose-Einstein condensates in interaction with thermal quasi-particles was addressed by a method  based on the equations of motion of the quasi-particle operators. This method leads directly to a quantum Langevin equation for the creation and annihilation operators of the collective modes, containing fluctuating force terms, a dissipation term, and a frequency shift term. These quantities are related by the fluctuation-dissipation relation and the Kramers-Kronig relation.
Each part of the interaction-Hamiltonian beyond the unperturbed Bogoliubov-Popov Hamiltonian in principle gives rise to separate contributions to all three types of terms. We have here considered only the most important of these, namely the part of the interaction Hamiltonian  giving rise to Landau-damping.

Dissipation can arise only from  energy conserving real processes, which is manifest by the appearance of the energy conserving $\delta$-functions in
the expressions for the damping rates. This means that only resonant processes can contribute to these rates.  In finite systems like the trapped condensates
this causes a problem, because there the mode spectrum is discrete, the spectrum of frequency differences $\omega_\kappa-\omega_\mu-\omega_\nu$ near $0$ is only
a quasi-continuum, and the dissipation rates in a strict sense have to vanish.
In other words, in a strict sense, what is seen as dissipation can only be
a 'short-time' effect; waiting for a sufficiently long time interval on the order of the inverse spacing of the quasi-continuum, revivals would have to appear. These will not be seen, however, at least in large condensates to which the local density and Thomas-Fermi approximation can be applied,
because not only the energy stored in the collective mode but also the thermal energy of the system is available to bring into play a large number of modes
which will lead to an irretrievable dissipation of the energy over many
degrees of freedom. Therefore it is reasonable in such cases, if not required, to eliminate
all recurrence effects, replacing the quasi-continuum by a true continuum, which is what  the local density approximation does.
Using this device we have arrived at definite results for the temperature-dependent damping rates  of any collective mode,
in an isotropic trap, which can be evaluated by computing numerically a multi-dimensional definite integral, e.g. by a Monte-Carlo routine.

The different pieces of the perturbation Hamiltonian also each give rise to frequency shifts. These are generated by virtual processes which do not require energy conservation, i.e. resonance. However the effect of the non-resonant
processes is suppressed by corresponding energy-denominators and small.
Here we
have limited our considerations only to those processes which can also
become resonant. We have here not  evaluated the frequency shifts further
using the local density approximation as we have done for the damping rates.

Experimental results for temperature dependent damping rates and frequency shifts have been obtained for {\it anisotropic} traps only \cite{rev1}, and we therefore refrain from a comparison with our explicit results for isotropic traps. Detailed comparisons have been made for anisotropic
traps in \cite{reidl2}, \cite{gior} where heavier and more powerful formalisms were brought to bear together with a stronger reliance on numerical work.

The goal here has been more modest, namely to use
 a minimum amount of numerical work and to apply the direct and intuitive
 quantum Langevin approach to the fluctuations, damping rates and frequency shifts of collective modes in spatially inhomogeneous
trapped Bose-Einstein condensates.

\section*{Acknowledgment:}

This work has been supported by the Deutsche Forschungsgemeinschaft through
the Sonderforschungsbereich 237 ``Unordnung und gro{\ss}e Fluktuationen''.
I wish to thank Martin Fliesser for useful discussions and owe thanks
to J\"urgen Reidl for the numerical evaluation of some
integrals.

\end{document}